\documentclass[reprint,showpacs,aps,pre,preprintnumbers]{revtex4-1}
\usepackage{graphicx}
\begin{document}

\title{Stochastic heat engine with the consideration of inertial effects and shortcuts to adiabaticity}
\author{Z. C. Tu}\email[Email: ]{tuzc@bnu.edu.cn}
\affiliation{Department of Physics, Beijing Normal University, Beijing 100875, China}
\affiliation{Beijing Computational Science Research Center, Beijing 100084, China}

\date{\today}

\begin{abstract}When a Brownian particle in contact with a heat bath at a constant temperature is controlled by a time-dependent harmonic potential, its distribution function can be rigorously derived from the Kramers equation with the consideration of the inertial effect of the Brownian particle. Based on this rigorous solution and the concept of shortcuts to adiabaticity, we construct a stochastic heat engine by employing the time-dependent harmonic potential to manipulate the Brownian particle to complete a thermodynamic cycle. We find that the efficiency at maximum power of this stochastic heat engine is equal to $1-\sqrt{T_c/T_h}$ where $T_c$ and $T_h$ are the temperatures of the cold bath and the hot one in the thermodynamic cycle, respectively.
\end{abstract}
\pacs{05.70.Ln, 05.40.Jc}
\preprint{Phys. Rev. E 89, 052148 (2014) \hspace{3cm}}
\maketitle

\small

\section{Introduction}

The concept of heat engines is a classical subject in thermodynamics. To achieve the highest efficiency, a heat engine needs to operate a reversible thermodynamic cycle which requires at least a quasi-static process~\cite{Thomsen60,Bizarro12} and results in a vanishing power. The thermodynamic cycle should be speeded up to produce a finite power. In our times of energy shortage, it is valuable to investigate how large the efficiency of a heat engine can be reached when the engine operates in the region of maximum power. This issue has lead to the birth of finite-time thermodynamics which has attracted much attention~\cite{Chambadal,Novikov,Curzon1975,Andresen77,Hoffmann85,devos85,Chen1989,ChenJC94,Bejan96,vdbrk2005,dcisbj2007,Esposito2010,GaveauPRL10,WangTu2011,wangtu2012,Izumida2012,wangheinter,Guochen2013,Apertet12a} for many years. The most notable result in finite-time thermodynamics is the Curzon-Ahlborn efficiency, $\eta_{CA}\equiv 1-\sqrt{T_c/T_h}$, which is the efficiency at maximum power for a macroscopically endoreversible heat engine~\cite{Curzon1975} operating between a cold bath at temperature $T_c$ and a hot bath at temperature $T_h$.

Recently, researchers started discussing the availability of the Curzon-Ahlborn efficiency for the microscopic models~\cite{Schmiedl2008,Tu2008,Esposito2009a} of heat engines. Schmiedl and Seifert~\cite{Schmiedl2008} constructed a stochastic heat engine by using a time-dependent harmonic potential to control a Brownian particle. Within the framework of stochastic thermodynamics~\cite{Sekimoto97,Udo08EPJB,Seifert12rev}, they fully investigated the energetics of this engine without the consideration of inertial effects. They found that the efficiency at maximum power of this stochastic heat engine is smaller than the Curzon-Ahlborn efficiency. The present author~\cite{Tu2008} investigated the energetics of the Feynman ratchet as a heat engine and found that the efficiency at maximum power of the Feynman ratchet is larger than the Curzon-Ahlborn efficiency. Esposito \emph{et al.}~\cite{Esposito2009a} found that the efficiency at maximum power of a quantum-dot heat engine is even larger than that of Feynman ratchet. On the one hand, three kinds of microscopic models mentioned above hint a universality of efficiency at maximum power up to the quadratic order for heat engines operating between two heat baths at small temperature difference. This universality has been confirmed by a model of particle transport~\cite{Esposito2009,Espositopre12}, a quantum-dot engine~\cite{Esposito41106}, and a generic model~\cite{ShengTuJPA13,ShengTuPRE13} of heat engines. On the other hand, these researches imply that it is quite difficult to construct a microscopic model of heat engines which can exactly achieve the Curzon-Ahlborn efficiency, $\eta_{CA}\equiv 1-\sqrt{T_c/T_h}$.

Although great progress has been made in finite-time thermodynamics, one of key challenges is still the realizability of finite-time adiabatic processes. Generally speaking, the adiabaticity requires a slowly enough process since the irreversibility usually accompanies a finite-time process. If finite-time adiabatic processes are proved to be impossible, the validation of the Curzon-Ahlborn efficiency and many main results in finite-time thermodynamics is questionable. Therefore, it is extremely urgent for us to solve this challenge. The concept of shortcuts to adiabaticity~\cite{Berry2009,ChenPRL10,Jarzynski40101,Campo100502,delCampoAX13,DengJ4182,DJdCAX14} developed in recent years throws light on this challenge. By utilizing shortcuts to adiabaticity in a quantum thermodynamical cycle, del Campo \emph{et al.} constructed an Otto heat engine working at finite power and zero friction~\cite{delCampoAX13}. Deng \emph{et al.} also found that the use of shortcuts to adiabaticity can increase the efficiency and the power of Otto heat engines in both quantum and classical regimes~\cite{DengJ4182}. In this paper, we will construct a solvable model of stochastic heat engines following the work by Schmiedl and Seifert~\cite{Schmiedl2008} with the consideration of the inertial effect of the Brownian particle and shortcuts to adiabaticity. Surprisingly, the efficiency at maximum power of this microscopically stochastic model is found to be exactly equal to the Curzon-Ahlborn efficiency. The rest of this paper is organized as follows. In Sec.~\ref{sec-stotherm}, we discuss the stochastic thermodynamics based on the Kramers equation (or the underdamping Fokker-Planck equation~\cite{vanKampen}), which will be applied to the investigation on the finite-time ``isothermal" processes in a thermodynamic cycle. In Sec.~\ref{sec-shtadiab}, we describe the concept of shortcuts to adiabaticity via a harmonic oscillator, which is helpful to investigate the finite-time ``adiabatic" processes in the thermodynamic cycle. In Sec.~\ref{sec-model}, we construct a  thermodynamic cycle by using a time-dependent harmonic potential to control the Brownian particle. In Sec.~\ref{sec-energ}, we investigate the energy transaction and the entropy variation in the thermodynamic cycle. In Sec.~\ref{sec-emp}, we discuss the efficiency at maximum power of our stochastic heat engine. The last section is a brief summary.

\section{Stochastic thermodynamics\label{sec-stotherm}}
Based on the Kramers equation, we generalize the framework of stochastic thermodynamics developed by Sekimoto~\cite{Sekimoto97} and Seifert~\cite{Udo08EPJB,Seifert12rev} in this section. The aim of this section is to lay the foundation for the discussions on energetics of ``isothermal" processes in the thermodynamic cycle introduced in Sec.~~\ref{sec-model}.

\subsection{General framework}
Let us consider a 1-dimensional movement of a Brownian particle in a heat bath at temperature $T$. A time-dependent potential $U(x,\lambda(t))$ is applied on the particle, where $x$ is the spatial coordinate of the particle while the function $\lambda=\lambda(t)$ represents the controlled protocol. Let us take $t$ and $p$ as the time variable and the momentum of the particle, respectively. Both the mass of particle and the Boltzmann constant are set to 1 in the present paper. The equation of motion may be expressed as the Langevin equation~\cite{vanKampen}:
\begin{equation}\label{eq-Langevin}
    \dot{x}=p,~\dot{p} =-\frac{\partial U(x,\lambda (t))}{\partial x}-\gamma p+\zeta (t),
\end{equation}
where $\gamma$ is the damping constant while $\zeta (t)$ represents Gaussian white noise satisfying $\langle\zeta (t) \rangle=0$ and $\langle\zeta (t)\zeta (0) \rangle=2\gamma T\delta(t)$. In this paper, the dot on a variable represents the total derivative of that variable with respect to time.
In the overdamping case, the inertial effect of the particle can be neglected. The stochastic thermodynamics without the consideration of inertial effects has been fully investigated by Sekimoto~\cite{Sekimoto97} and Seifert~\cite{Udo08EPJB,Seifert12rev}. It is straightforward to extend their thoughts into the underdamping case where the inertial effect of the particle plays a substantial role. The Hamiltonian of the particle may be expressed as \begin{equation}H=\frac{p^{2}}{2}+U(x,\lambda (t)).\label{eq-Hamtion0}\end{equation} The differential of the Hamiltonian can be expressed as
\begin{equation}\mathrm{d}H=\left[\dot{p}p+\dot{x}\frac{\partial U}{\partial x }\right]\mathrm{d}t+\left[\dot{\lambda}\frac{\partial U}{\partial\lambda }\right] \mathrm{d}t,\end{equation}
which enlightens us to define the energy difference
\begin{equation}\Delta e\equiv H(t_f)-H(t_i),\label{eq-detae}\end{equation}
the input work~\cite{JarzynskiPRL97,JarzynskiPRX13}
\begin{equation}w\equiv \int_{t_i}^{t_f}\mathrm{d}t\dot{\lambda}\frac{\partial U}{\partial\lambda },\label{eq-detaw}\end{equation}
and the absorbed heat
\begin{equation}q\equiv \int_{t_i}^{t_f}\mathrm{d}t\left[\dot{p}p+\dot{x}\frac{\partial U}{\partial x }\right]\label{eq-detaq}\end{equation}
along a phase trajectory $\{x(t),p(t)\}$ stemming from a phase point $(x_i, p_i)$ at initial time $t_i$ and ending at a phase point $(x_f, p_f)$ at final time $t_f$.
The energy balance
\begin{equation}
\Delta e=w+q\end{equation}
holds for each phase trajectory.

Corresponding to the Langevin equation~(\ref{eq-Langevin}), the distribution function $\rho(x,p,t)$ of the particle is governed by the Kramers equation~\cite{vanKampen,Seifert12rev}:
\begin{equation}\label{eq-Kramers}
    \frac{\partial \rho}{\partial t}+\nabla\cdot \mathbf{J}=0
\end{equation}
with flux
\begin{equation}
\mathbf{J}\equiv p\rho \hat{\mathbf{x}}-\rho\left( \gamma p+\frac{\partial U}{\partial x}+\frac{\gamma T}{\rho}\frac{%
\partial \rho}{\partial p}\right) \hat{\mathbf{p}}\label{eq-Kramersflux}
\end{equation}
and gradient operator $\nabla\equiv \hat{\mathbf{x}}{\partial}/{\partial x}+\hat{\mathbf{p}}{\partial}/{\partial p}$,
where $\hat{\mathbf{x}}$ and $\hat{\mathbf{p}}$ represent the unit vectors in the coordinates of position and momentum of the particle.

The ensemble averages of the quantities in Eqs.(\ref{eq-detae})-(\ref{eq-detaq}) can be calculated via the similar procedure in Refs.~\cite{Udo08EPJB,ShizumePRE95,BizarroPRE11}. The average energy difference and the average input work may be expressed as
\begin{equation}\Delta E\equiv \langle\Delta e\rangle= \left.\int \mathrm{d}x\int \mathrm{d}p(H\rho)\right|_{t_{i}}^{t_{f}},\label{eq-detaee}\end{equation}
and
\begin{equation}W\equiv \langle w \rangle= \int_{t_{i}}^{t_{f}}\mathrm{d}t\int \mathrm{d}x\int \mathrm{d}p \left(\rho \dot{\lambda}\frac{\partial U}{\partial\lambda }\right).\label{eq-detaew}\end{equation}
Then using the energy balance and the Kramers equation, we may derive the average heat absorbed from the medium:
\begin{eqnarray} Q&\equiv &\langle q \rangle= \int_{t_{i}}^{t_{f}}\mathrm{d}t\int \mathrm{d}x\int \mathrm{d}p (\mathbf{J}\cdot \nabla H )\nonumber\\ &=& - \int_{t_{i}}^{t_{f}}\mathrm{d}t\int \mathrm{d}x\int \mathrm{d}p\left[ \gamma p\rho\left( p+\frac{T}{\rho}%
\frac{\partial \rho}{\partial p}\right) \right]
.\label{eq-detaeq}\end{eqnarray}
The detailed derivation of the above equation is attached in Appendix~\ref{sec-Apdx1}.

In addition, the ensemble average of trajectory entropy may be defined as~\cite{Udo08EPJB,Seifert12rev}
\begin{equation}S\equiv \langle -\ln \rho\rangle=-\int \mathrm{d}x\int\mathrm{d}p(\rho\ln \rho)\label{eq-entropy}.\end{equation}
By considering this definition and the Kramers equation, we may derive the variation of entropy
\begin{equation}\Delta S=\int_{t_{i}}^{t_{f}}\mathrm{d}t\int \mathrm{d}x\int \mathrm{d}p \left[\gamma \frac{\partial \rho}{\partial p}\left( p+\frac{ T}{\rho}
\frac{\partial \rho}{\partial p}\right)\right]\label{eq-deltaS}.\end{equation}
Thus the energy dissipation $R\equiv T\Delta S - Q$ may be expressed as
\begin{equation}R =\int_{t_{i}}^{t_{f}}\mathrm{d}t\int \mathrm{d}x\int \mathrm{d}p\left[\gamma \rho\left( p+\frac{T}{\rho}\frac{\partial \rho}{\partial p}\right) ^{2}\right]\ge 0\label{eq-dissip}.\end{equation}

\subsection{Pedagogical example: Brownian particle in a time-dependent harmonic potential}

Now we consider a Brownian particle in a time-dependent harmonic potential $U=\lambda^2(t) x^2/2$. Its Hamiltonian can be expressed as
\begin{equation}\label{eq-Hharm}
    H(t)=\frac{p^{2}}{2}+\frac{\lambda^2(t) x^2}{2}.\end{equation}
It is not hard to verify that the distribution function
\begin{equation}\label{eq-pdf0}
    \rho =\frac{\beta (t)\lambda (t)}{2\pi }\exp \left[ -\beta(t) \left( \frac{p^{2}
}{2}+\frac{\lambda ^{2}(t)x^{2}}{2}\right) \right]
\end{equation}
is a special solution to the Kramers equation (\ref{eq-Kramers}) if
\begin{equation}\beta(t) \lambda^{2}(t)=\mathrm{constant}\label{eq-condt1}\end{equation}
and
\begin{equation}\frac{\mathrm{d} \beta(t)}{\mathrm{d}t}=2\gamma
\beta(t) [ 1-T\beta(t) ]\label{eq-condt2}\end{equation}
are simultaneously satisfied. This is our first key result in the present paper.

The above distribution function (\ref{eq-pdf0}) implies that
\begin{equation}\label{eq-equienerg}
    \left\langle \frac{p^2}{2}\right\rangle =\left \langle \frac{\lambda^{2}(t) x^2}{2}\right\rangle=\frac{1}{2\beta(t)},
\end{equation}
which may be regarded as the equipartition of energy if we interpret $1/\beta(t)$ as the effective temperature of the ensemble of Brownian particles in the time-dependent potential. By combining Eqs.~(\ref{eq-condt1}) and (\ref{eq-equienerg}), we derive $\langle x^2 \rangle= \mathrm{constant}$ and $\langle p^2 \rangle \propto \lambda ^{2}(t)$, which imply that the width of the position distribution is time-independent while the width of the momentum distribution is expanded when the potential is enhanced ($\lambda$ increases with time), and vice versa.

With the consideration of Hamiltonian (\ref{eq-Hharm}) and distribution function (\ref{eq-pdf0}), Eqs.~(\ref{eq-detaee}), (\ref{eq-detaeq}), (\ref{eq-deltaS}) and (\ref{eq-dissip}) may be transformed into
\begin{equation}\label{eq-delteehm}
\Delta E={1/\beta(t_{f})}-{1/\beta(t_{i})},
\end{equation}
\begin{equation}\label{eq-delteqhm}
Q=-\gamma \int_{t_{i}}^{t_{f}}\mathrm{d}t [1/\beta(t)- T],
\end{equation}
\begin{equation}\label{eq-entrham}
\Delta S=({1}/{2})\ln [\beta(t_i)/\beta(t_f)],
\end{equation}
and
\begin{equation}\label{eq-dispham}
R=\gamma \int_{t_{i}}^{t_{f}}\mathrm{d}t [1-\beta(t) T] ^{2}/\beta(t),
\end{equation}
respectively.
The energy dissipation $R$ is nonnegative and it vanishes merely for the equilibrium state $\beta(t)=1/T=\mathrm{constant}$.

\section{Shortcuts to adiabaticity\label{sec-shtadiab}}
Researchers have always thought that the realization of an adiabatic change requires an extremely slow control to the system. For example, the area of phase space enclosed in an energy shell for a one-dimensional system in classical mechanics may be expressed as $I=\oint p \mathrm{d}x$ where the integral is taken over the path in the phase space for a given energy and driving protocol. Classical mechanics tells us that the quantity $I$ is an adiabatic invariant, remaining constant along a Hamiltonian trajectory $\{x(t),p(t)\}$ when the protocol is varied infinitely slowly~\cite{Landaubook}. In quantum mechanics, the adiabatic theorem~\cite{BornFock28,Berry2009} implies that a physical system remains in its instantaneous eigenstate if a given perturbation is varied slowly enough and if there is a gap between the corresponding eigenvalue and the others.
Recently, it was found that it is possible to generate a shortcut to adiabaticity under which the value of classical quantity $I$ is preserved exactly, and the quantum system remains in its instantaneous eigenstate even the driving protocol is varied in a finite rate~\cite{DengJ4182,Jarzynski40101,Campo100502,delCampoAX13,DJdCAX14}. For simplicity, we consider a Brownian particle in a time-dependent harmonic potential again. The Hamiltonian is still expressed as Eq.~(\ref{eq-Hharm}). During time $t_i<t\le t_f$, the protocol varies from $\lambda_i\equiv\lambda (t_i)$ to $\lambda_f\equiv\lambda (t_f)$. With consideration of a counterdiabatic
driving Hamiltonian
\begin{equation}\label{eq-coHharm}
    H_C(t)=H(t)-\frac{\dot{\lambda}(t)}{2\lambda(t)}xp,\end{equation}
with $H(t)$ being the original system Hamiltonian~(\ref{eq-Hharm}), the evolution of the system can be enforced along the adiabatic manifold of the system Hamiltonian~\cite{DengJ4182,Jarzynski40101,Campo100502,delCampoAX13,DJdCAX14}. The only requirement is $\dot{\lambda}(t_i)=\dot{\lambda}(t_f)=0$ such that $H_C(t_i)=H(t_i)$ at the initial time  $t_i$ and $H_C(t_f)=H(t_f)$ at the final time $t_f$.
There exists a certain arbitrariness for selecting the protocol $\lambda (t)$. One simple choice is
\begin{equation}\label{eq-protadbat}\lambda(t)=\lambda_i + (\lambda_f-\lambda_i)\Phi\left(\frac{t-t_i}{t_f-t_i}\right),\end{equation}
where the function $\Phi(t)$ is defined as $\Phi(t)\equiv 3 t^{2}-2t^{3}$. Note that the main results in the present paper are independent of this choice.

The equations of motion governed by the counterdiabatic
driving Hamiltonian $H_C(t)$ may be expressed as
\begin{equation}\label{eq-Hameq}
\left\{\begin{array}{l}
        \dot{x}=\frac{\partial H_C}{\partial p}=p-\frac{\dot{\lambda}(t)}{2\lambda(t)}x \\
         \dot{p}=-\frac{\partial H_C}{\partial x}=-\lambda^2(t) x+\frac{\dot{\lambda}(t)}{2\lambda(t)}p
       \end{array}.
 \right.
\end{equation}
According to the above equations of motion, it is not hard to verify that the value of $I=\oint p \mathrm{d}x \propto H(t)/\lambda(t)$ is conserved exactly along the Hamiltonian trajectory $\{x(t),p(t)\}$, for any protocol $\lambda(t)$~\cite{Jarzynski40101}.

In the following discussion, we will prove that the shortcut to adiabaticity
can link two canonical states with different effective temperatures. This is our second key result in the present paper which is crucial to construct the ``adiabatic" processes in the thermodynamic cycle in Sec.~\ref{sec-model}.

Assume that the system initially stays in a canonical state with effective temperature $\beta_i^{-1}$. The initial distribution function of the system may be expressed as
\begin{equation}\label{eq-pdfin}
    \rho_i = \frac{\beta_i \lambda_i}{2\pi}\exp[-\beta_i H(t_i)]
\end{equation}
with Hamiltonian
\begin{equation}\label{eq-Hamti}
H(t_i)=p_i^2/2+\lambda^2_ix_i^2/2,\end{equation}
where $(x_i,p_i)$ represents the point in the phase space at initial time $t_i$.
According to the Liouville theorem, the
distribution function is invariant along the phase trajectory since the microscopic motions abide by the Hamilton equation (\ref{eq-Hameq}) when the system is not in contact with any heat bath, that is, the distribution function of final state should be
\begin{equation}\label{eq-pdffn}
    \rho_f =\rho_i =\frac{\beta_i \lambda_i}{2\pi}\exp[-\beta_i H(t_i)].
\end{equation}

We will seek an effective temperature $\beta_f^{-1}$ such that the distribution function (\ref{eq-pdfin}) may be expressed as a canonical distribution
\begin{equation}\label{eq-pdffn1}
    \rho_f = \frac{\beta_f \lambda_f}{2\pi}\exp[-\beta_f H(t_f)]
\end{equation}
with Hamiltonian
\begin{equation}\label{eq-Hamtf}
H(t_f)=p_f^2/2+\lambda^2_fx_f^2/2,\end{equation}
where $(x_f,p_f)$ represents the point in the phase space at final time $t_f$.
According to the Hamilton equation (\ref{eq-Hameq}), we can derive
\begin{equation}\frac{\mathrm{d}H(t)}{\mathrm{d}t}=H(t)\frac{\mathrm{d}\ln\lambda(t)}{\mathrm{d}t},\label{eq-dHdt}\end{equation}
which leads to $H(t_f)=H(t_i)\lambda_f/\lambda_i$. The derivation of the above equation is attached in Appendix~\ref{sec-Apdx2}. By substituting this equation into Eq.~(\ref{eq-pdffn1}), we obtain
\begin{equation}\label{eq-pdffn2}
    \rho_f = \frac{\beta_f \lambda_f}{2\pi}\exp\left[-\frac{\beta_f \lambda_f}{\lambda_i} H(t_i)\right].
\end{equation}
By comparing the above equation with Eq.~(\ref{eq-pdffn}), we obtain
\begin{equation}\beta_f\lambda_f=\beta_i \lambda_i,\label{eq-effctemp}\end{equation}
which implies that the system will stay finally in the canonical state with effective temperature $\beta_f^{-1}$ after it undergoes the shortcut to adiabaticity governed by the Hamiltonian equation (\ref{eq-Hameq}) if it initially stays in the canonical state with effective temperature $\beta_i^{-1}$. The effective temperatures of the initial state and the final state should satisfy Eq.~(\ref{eq-effctemp}). Since the microscopic motion abides by the Hamilton equation (\ref{eq-Hameq}), there is no heat exchange and entropy production in the shortcut to adiabaticity during time interval $t_f-t_i$. Therefore, we can realize the relatively quick but adiabatic transition from one canonical state to another compatible canonical state. In addition, since both the initial state and the final state are canonical, the width of momentum distribution and the energy difference between these two canonical states still satisfy Eqs.~(\ref{eq-equienerg}) and (\ref{eq-delteehm}), respectively.

\section{Model\label{sec-model}}
\begin{figure}[htp!]
\includegraphics[width=7cm]{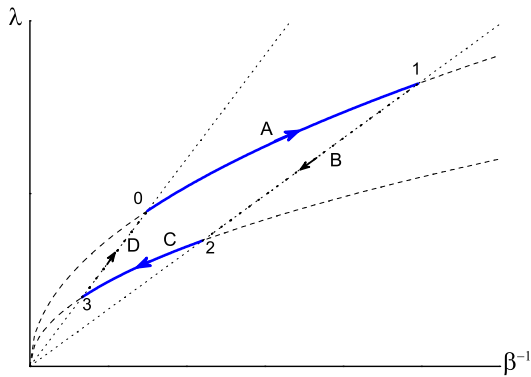}\caption{(Color online) Thermodynamic cycle. The dashed lines correspond to Eq.~(\ref{eq-condt1}) while the dotted lines correspond to $\beta(t)\lambda(t)=\mathrm{constant}$, the continuation of Eq.~(\ref{eq-effctemp}). The capital letters A, B, C, and D represent four processes in the thermodynamic cycle, respectively.}\label{fig-cycle}
\end{figure}

We construct a Carnot-like thermodynamic cycle by using a time-dependent harmonic potential  $U=\lambda^2(t) x^2/2$ to manipulate a Brownian particle. As depicted in Fig.~\ref{fig-cycle}, the cycle consists of four processes as follows.

\subsection{``Isothermal" expansion}
This process corresponds to the solid line linking 0 and 1 in Fig.~\ref{fig-cycle}. Here the word ``isothermal" merely indicates
that the Brownian particle is in contact with a hot bath at constant temperature $T_h$. It does not mean that the effective temperature of the system is constant. During time $0<t\le t_1$, the protocol $\lambda(t)$ varies monotonically from $\lambda(0)\equiv\lambda_0$ to $\lambda(t_1)\equiv\lambda_1(>\lambda_0)$. According to the discussion below Eq.~(\ref{eq-equienerg}), the width of momentum distribution is expanded when $\lambda$ increases with time. It is in this sense that this process is referred to as an expansion.

On the other hand, Eq.~(\ref{eq-condt1}) implies that $\beta(t)$ decreases with time when $\lambda(t)$ increases with time. Then from Eq.~(\ref{eq-condt2}) we solve
\begin{equation}\label{eq-betaivsA}
 1/\beta(t) =T_{h}( 1-c_{h}\mathrm{e}^{ -2\gamma_h t}),
\end{equation}
with a parameter $c_h>0$. $\gamma_h$ represents the damping constant for the particle in the hot bath.
From Eq.~(\ref{eq-condt1}) we obtain the protocol
\begin{equation}\label{eq-protA}
 \lambda(t) =\lambda _{0}\sqrt{(1-c_{h}\mathrm{e}^{
-2\gamma_h t} )/(1-c_{h})}.
\end{equation}
In particular, we have
\begin{equation}\label{eq-beta0}
 \beta_0^{-1}\equiv 1/\beta(0)=T_{h}(1-c_{h}),
\end{equation}
\begin{equation}\label{eq-beta1}
  \beta_1^{-1}\equiv 1/\beta(t_1)=T_{h}( 1-c_{h}\tau_h),
\end{equation}
and
\begin{equation}\label{eq-lambda1}
 \lambda_1\equiv \lambda(t_1) =\lambda _{0}\sqrt{(1-c_{h} \tau_h )/(1-c_{h})},
\end{equation}
where $\tau_h\equiv \mathrm{e}^{ -2\gamma_h t_1} $.

\subsection{``Adiabatic" compression}
This process corresponds to the dotted line linking 1 and 2 in Fig.~\ref{fig-cycle}.
With the aid of shortcuts to adiabaticity discussed in Sec.~\ref{sec-shtadiab}, the protocol $\lambda(t)$ varies monotonically from $\lambda(t_1^+)\equiv\lambda_1$ to $\lambda(t_2)\equiv \lambda_2(<\lambda_1)$ during time $t_1<t\le t_2$. The whole system is not in contact with any heat bath. According to Eq.~(\ref{eq-effctemp}), the effective temperature $\beta_2^{-1}$ at time $t_2$ should satisfy
\begin{equation}\beta_2\lambda_2=\beta_1\lambda_1.\label{eq-beta2}\end{equation}

Note that this process is marked with the dotted line in Fig.~\ref{fig-cycle} because we cannot actually define the effective temperature of the system in the whole process except at times $t_1$ and $t_2$. In fact, it is unnecessary for us to define the effective temperature in this precess except at times $t_1$ and $t_2$. On the other hand, we find $\beta_2^{-1}<\beta_1^{-1}$ from Eq.~(\ref{eq-beta2}) since $\lambda_2<\lambda_1$. According to the discussion in Sec.~\ref{sec-shtadiab}, the states at times $t_1$ and $t_2$ are two canonical states with effective temperatures $\beta_1^{-1}$ and $\beta_2^{-1}$, respectively. Thus the width of momentum distribution at time $t_2$ is narrower than that at time $t_1$ with the consideration of Eq.~(\ref{eq-equienerg}). It is in this sense that we call this process an ``adiabatic" compression.

\subsection{``Isothermal" compression}
This process corresponds to the solid line linking 2 and 3 in Fig.~\ref{fig-cycle}. Here the word ``isothermal" merely indicates
that the Brownian particle is in contact with a cold bath at constant temperature $T_c$. During time $t_2<t\le t_3$, the protocol $\lambda(t)$ varies monotonically from $\lambda(t_2^+)\equiv\lambda_2$ to $\lambda(t_3)\equiv\lambda_3(<\lambda_2)$. According to the discussion below Eq.~(\ref{eq-equienerg}), the width of momentum distribution narrows down when $\lambda$ decreases with time. It is in this sense that this process is called a compression.

On the other hand, Eq.~(\ref{eq-condt1}) implies that $\beta(t)$ increases with time when $\lambda(t)$ decreases with time. Then from Eq.~(\ref{eq-condt2}) we solve
\begin{equation}\label{eq-betaivsC}
 1/\beta(t) =T_{c}[1+c_{c}\mathrm{e}^{ -2\gamma_c (t-t_{2})} ],
\end{equation}
with a parameter $c_c>0$. $\gamma_c$ represents the damping constant for the particle in the cold bath. The effective temperature at time $t_2$ may be expressed as
\begin{equation}\label{eq-beta22}
 \beta_2^{-1}\equiv 1/\beta (t_2)=T_{c}(1+c_{c}).
\end{equation}
By considering Eqs.~(\ref{eq-beta1})-(\ref{eq-beta2}) and (\ref{eq-beta22}), we obtain
\begin{equation}\label{eq-lambda22}
\lambda_{2}=\frac{T_{c}( 1+c_{c}) }{T_{h}\sqrt{1-c_{h}} }\frac{\lambda
_{0}}{\sqrt{{1-c_{h}\tau_h }}}.
\end{equation}
With the consideration of Eq.~(\ref{eq-condt1}), we obtain the protocol
\begin{equation}\label{eq-protC}
 \lambda(t) =\lambda _{2}\sqrt{[
1+c_{c}\mathrm{e}^{ -2\gamma_c ( t-t_{2}) } ]/(
1+c_{c})} .
\end{equation}
In particular, from the above equation and Eq.~(\ref{eq-betaivsC}) we have
\begin{equation}\label{eq-beta3}
  \beta_3^{-1}\equiv 1/\beta(t_3)=T_{c}(1+c_{c}\tau_c),
\end{equation}
and
\begin{equation}\label{eq-lambda3}
 \lambda_3\equiv \lambda(t_3) =\lambda _{0}\frac{T_{c}\sqrt{1+c_{c}}}{T_{h}
\sqrt{1-c_{h}}}\sqrt{\frac{1+c_{c}\tau_c }{1-c_{h}\tau_h }}
\end{equation}
where $\tau_c\equiv\mathrm{e}^{ -2\gamma_c
(t_{3}-t_2)}$.

\subsection{``Adiabatic" expansion}
This process corresponds to the dotted line linking 3 and 0 in Fig.~\ref{fig-cycle}.
With the aid of shortcuts to adiabaticity discussed in Sec.~\ref{sec-shtadiab}, the protocol $\lambda(t)$ varies monotonically from $\lambda(t_3^+)\equiv\lambda_3$ to $\lambda(t_4)\equiv \lambda_4(>\lambda_3)$ during time $t_3<t\le t_4$. The whole system is not in contact with any heat bath. According to Eq.~(\ref{eq-effctemp}), the effective temperature $\beta_4^{-1}$ at time $t_4$ should satisfy
\begin{equation}\beta_4\lambda_4=\beta_3\lambda_3.\label{eq-beta4}\end{equation}

Note that this process is marked with the dotted line in Fig.~\ref{fig-cycle} because we cannot define the effective temperature of the system in the whole process except at times $t_3$ and $t_4$. In fact, it is unnecessary for us to do that except at times $t_3$ and $t_4$. On the other hand, we find $\beta_4^{-1}>\beta_3^{-1}$ from Eq.~(\ref{eq-beta4}) since $\lambda_4>\lambda_3$. According to the discussion in Sec.~\ref{sec-shtadiab}, the states at times $t_3$ and $t_4$ are two canonical states with effective temperatures $\beta_3^{-1}$ and $\beta_4^{-1}$, respectively. Thus the width of momentum distribution at time $t_4$ is wider than that at time $t_3$ with the consideration of Eq.~(\ref{eq-equienerg}). It is in this sense we call this process an ``adiabatic" expansion.

To make a full cycle, the periodic conditions $\lambda _{4}=\lambda _{0}$ and $\beta _{4}=\beta _{0}$ should be imposed, which lead to a constraint
\begin{equation}\label{eq-constrat}
( 1+c_{c})( 1-c_{h}) =(1-c_{h}\tau_h)( 1+c_{c}\tau_c)
\end{equation}
with the consideration of Eqs.~(\ref{eq-beta0}) and (\ref{eq-beta3})--(\ref{eq-beta4}).

It should be emphasized that we have explicitly constructed a new type of thermodynamic cycle
which is different from that considered by Curzon and Ahlborn. The effective temperature of the ``isothermal" processes is assumed to be constant in the Curzon-Ahlborn model~\cite{Curzon1975}. It is still unclear whether the Curzon-Ahlborn model is reliable within the framework of statistical mechanics.
While in the present model, the relation between the value of protocol $\lambda (t)$ and the time-dependently effective temperature $1/\beta(t)$ of the ``isothermal" processes is well-defined. In this sense, Fig.~\ref{fig-cycle} may be regarded as a counterpart of $PT$-diagram of a reversible engine in the present irreversible model. This new construction is our third key contribution in the present paper.

\section{Energetics\label{sec-energ}}
In this section, we will investigate the energy transaction and the entropy variation in the four processes mentioned above.

First, in the ``isothermal" expansion, we obtain the energy difference
\begin{equation}\label{eq-delteisoh}
\Delta E_A\equiv \beta_1^{-1}-\beta_0^{-1}=T_hc_h(1-\tau_h),
\end{equation}
the heat absorbed from the hot bath
\begin{equation}\label{eq-deltqisoh}
Q_A\equiv-\gamma_h \int_{0}^{t_{1}}\mathrm{d}t [1/\beta(t)- T_h]=\frac{1}{2}T_hc_h(1-\tau_h),
\end{equation}
the work input
\begin{equation}\label{eq-workisoh}
W_A\equiv \Delta E_A-Q_A=\frac{1}{2}T_hc_h(1-\tau_h),
\end{equation}
the entropy variation
\begin{equation}\label{eq-entisoh}
\Delta S_A\equiv\frac{1}{2}\ln \frac{\beta_0}{\beta_1}=\frac{1}{2}\ln \frac{1-c_{h}\tau_h}{1-c_{h}}\ge 0,
\end{equation}
and the energy dissipation
\begin{eqnarray}\label{eq-dispisoh}
R_A &\equiv& \gamma_h \int_{0}^{t_1}\mathrm{d}t [1-\beta(t) T_h] ^{2}/\beta(t)\nonumber\\ &=&\frac{T_{h}}{2}\left[\ln \frac{1-c_{h}\tau_h }{1-c_{h}}- c_{h}(1- \tau_h)\right]
\end{eqnarray}
from Eqs.~(\ref{eq-delteehm})-(\ref{eq-dispham}) and (\ref{eq-betaivsA})-(\ref{eq-beta1}).

Second, as mentioned below Eq.~(\ref{eq-effctemp}), since the system is not in contact with any heat bath and its evolution abides by the Hamilton equation (\ref{eq-Hameq}), both the heat exchange and the entropy production are vanishing in the ``adiabatic" compression, which are denoted by
\begin{equation}\label{eq-detaqentB}
Q_B=0~\mathrm{and}~\Delta S_B=0,
\end{equation}
respectively. The Hamilton equation is microscopically reversible, thus this ``adiabatic" compression is reversible in the level of dynamics. However, this ``adiabatic" compression is slightly different from a reversible process such as an adiabatic process in a conventional thermodynamic cycle. Here we merely indicate that the initial state 1 and the final state 2 in Fig.~\ref{fig-cycle} are located in an isentropic line. During the process of ``adiabatic" compression connecting these two states, the entropy and the temperature are not well-defined except at these two states, whereas, both the entropy and the temperature are well-defined in the whole adiabatic process in a conventional thermodynamic cycle. Similar discussion is also available for the ``adiabatic" expansion.
The work input and the energy difference may be expressed as
\begin{equation}\label{eq-workB}
W_B=\Delta E_B\equiv \beta_2^{-1}-\beta_1^{-1}=T_c(1+c_c)-T_h(1- c_h \tau_h)
\end{equation}
according to Eqs.~(\ref{eq-beta1}) and (\ref{eq-beta22}) as well as the discussion below Eq.~(\ref{eq-effctemp}).

Third, in the ``isothermal" compression, we obtain the energy difference
\begin{equation}\label{eq-delteisoc}
\Delta E_C\equiv \beta_3^{-1}-\beta_2^{-1}=-T_cc_c(1-\tau_c),
\end{equation}
the heat absorbed from the cold bath
\begin{equation}\label{eq-deltqisoc}
Q_C\equiv-\gamma_c \int_{t_2}^{t_3}\mathrm{d}t [1/\beta(t)- T_c]=-\frac{1}{2}T_cc_c(1-\tau_c),
\end{equation}
the work input
\begin{equation}\label{eq-workisoc}
W_C\equiv \Delta E_C-Q_C=-\frac{1}{2}T_cc_c(1-\tau_c),
\end{equation}
the entropy variation
\begin{equation}\label{eq-entisoc}
\Delta S_C\equiv\frac{1}{2}\ln \frac{\beta_2}{\beta_3}=\frac{1}{2}\ln \frac{1+c_{c}\tau_c}{1+c_{c}}\le 0,
\end{equation}
and the energy dissipation
\begin{eqnarray}\label{eq-dispisoc}
R_C &\equiv& \gamma_c \int_{t_2}^{t_3}\mathrm{d}t [1-\beta(t) T_c] ^{2}/\beta(t)\nonumber\\ &=&\frac{T_{c}}{2}\left[\ln \frac{1+c_{c}\tau_c}{1+c_{c}}+ c_{c}(1- \tau_c)\right]
\end{eqnarray}
from Eqs.~(\ref{eq-delteehm})-(\ref{eq-dispham}) and (\ref{eq-betaivsC})-(\ref{eq-beta3}).

Fourth, both the heat exchange and the entropy production are vanishing in the ``adiabatic" expansion, which are denoted by
\begin{equation}\label{eq-detaqentD}
Q_D=0~\mathrm{and}~\Delta S_D=0,
\end{equation}
respectively.
The work input and the energy difference may be expressed as
\begin{equation}\label{eq-workD}
W_D=\Delta E_D\equiv \beta_0^{-1}-\beta_3^{-1}=T_h(1-c_h)-T_c(1+ c_c \tau_c),
\end{equation}
according to Eqs.~(\ref{eq-beta0}) and (\ref{eq-beta3}) as well as the discussion below Eq.~(\ref{eq-effctemp}).

When the system completes a whole cycle, it will return to its initial state. Since the energy and the entropy of the system are state variables, both of them should be unchanged when the system completes the whole cycle. In fact, from Eqs.~(\ref{eq-delteisoh}), (\ref{eq-workB}), (\ref{eq-delteisoc}) and (\ref{eq-workD}), we can confirm $\Delta E_A+\Delta E_B+\Delta E_C+\Delta E_D =0$. With the consideration of Eqs.~(\ref{eq-entisoh}), (\ref{eq-detaqentB}), (\ref{eq-entisoc}), (\ref{eq-detaqentD}) and constraint (\ref{eq-constrat}), we can also verify $\Delta S_A+\Delta S_B+\Delta S_C+\Delta S_D =0$. The work output may be expressed as
\begin{eqnarray}
W_\mathrm{out}&\equiv&-(W_A+W_B+W_C+W_D)\nonumber\\
&=&\frac{1}{2}[T_hc_h(1-\tau_h)-T_cc_c(1-\tau_c)] \label{eq-workout}
\end{eqnarray}
from Eqs.~(\ref{eq-workisoh}), (\ref{eq-workB}), (\ref{eq-workisoc}) and (\ref{eq-workD}). This result is consistent with $W_\mathrm{out}= Q_A+Q_C$ directly derived from the energy balance in the whole cycle. When $W_\mathrm{out}>0$, this system operates as a heat engine.

The efficiency of the engine is defined as the ratio of the work output to the heat absorbed from the hot bath, which reads
\begin{equation}
\eta\equiv \frac{W_\mathrm{out}}{Q_A}=1-\frac{T_cc_c(1-\tau_c) }{T_hc_h(1-\tau_h)}
\label{eq-efficy}
\end{equation}
with the consideration of Eqs.~(\ref{eq-deltqisoh}) and (\ref{eq-workout}). It is not hard to verify that $\eta$ is less than the Carnot efficiency $\eta_C\equiv 1-T_c/T_h$ from Eq.~(\ref{eq-constrat}).

The power is defined as the work output divided by the period ($t_4$) for completing the whole cycle, which reads
\begin{equation}
P\equiv \frac{W_\mathrm{out}}{t_4}= \frac{T_hc_h(1-\tau_h)-T_cc_c(1-\tau_c) }{2t_4}.
\label{eq-powerout}
\end{equation}

\section{Efficiency at maximum power\label{sec-emp}}
Now let us optimize the heat engine. To maximize the power (\ref{eq-powerout}) under the constraint (\ref{eq-constrat}), we introduce a Lagrange multiplier $\Lambda$ and then seek the maximum of the extended function
\begin{eqnarray}
I&\equiv& \frac{T_hc_h(1-\tau_h)-T_cc_c(1-\tau_c) }{2t_4}\\&+&\Lambda [ ( 1-c_{h}\tau_h) ( 1+c_{c}\tau_c) -( 1+c_{c}) (1-c_{h}) ].\nonumber
\end{eqnarray}

The procedure of maximization is standard. From $\partial I/\partial c_h =0$ and $\partial I/\partial c_c =0$, we can obtain
\begin{equation}\label{eq-max1}
{T_h( 1-\tau_h)}/{2t_4}=\Lambda [( 1+c_{c}\tau_c)\tau_h -(1+c_{c})]
\end{equation}
and
\begin{equation}\label{eq-max2}
{T_c ( 1-\tau_c) }/{2t_4}=\Lambda [ ( 1-c_{h}\tau_h) \tau_c-( 1-c_{h})],
\end{equation}
respectively.
Divided Eq.~(\ref{eq-max1}) by Eq.~(\ref{eq-max2}), we arrive at
\begin{equation}\label{eq-max3}
\frac{T_h( 1-\tau_h)}{T_c ( 1-\tau_c) }=\frac{1-\tau_h+c_c(1-\tau_h\tau_c)}{1-\tau_c-c_h( 1-\tau_h\tau_c)}.
\end{equation}
On the other hand, from constraint equation (\ref{eq-constrat}), we have
\begin{equation}1-\tau_h\tau_c =(1-\tau_c)/c_{h} -(1-\tau_h)/c_{c}.\label{eq-constrt2}
\end{equation}
Substituting this equation into Eq.~(\ref{eq-max3}), we can obtain
\begin{equation}\label{eq-max4}
{c_c( 1-\tau_c) }/{c_h( 1-\tau_h)}=\sqrt{{T_h }/{T_c}}.
\end{equation}
Substituting the above equation into Eq.~(\ref{eq-efficy}), we obtain the efficiency at maximum power:
\begin{equation}\label{eq-CAefficy}
 \eta_{\mathrm{mP}}=1-\sqrt{{T_c }/{T_h}}.
\end{equation}
This is our fourth key result in the present paper. Interestingly, this result is exactly equal to the Curzon-Ahlborn efficiency for endoreversible heat engines working at maximum power although the present stochastic model looks quite different from Curzon-Ahlborn endoreversible heat engines. The effective temperatures are time-dependent and well-defined in the two ``isothermal" processes in our model while they are presumed to be constant in the Curzon-Ahlborn model. Furthermore, from Eqs.~(\ref{eq-constrt2}) and (\ref{eq-max4}) we can also solve:
\begin{equation}\label{eq-ch}
c_{h}=\left(1-\sqrt{{T_c }/{T_h}}\right)( 1-\tau_c)/(1-\tau_h\tau_c)
\end{equation}
and
\begin{equation}\label{eq-cc}
c_{c} =\left(\sqrt{{T_h }/{T_c}}-1\right)(1-\tau_h)/(1-\tau_h\tau_c).
\end{equation}
By substituting the above two equations into Eq.~(\ref{eq-powerout}), we achieve the maximum power
\begin{equation}\label{eq-powermaxm}
P_\mathrm{max}\propto \left(\sqrt{T_h}-\sqrt{T_c }\right)^2,
\end{equation}
which displays the same behavior as the maximum power of Curzon-Ahlborn endoreversible heat engines~\cite{Curzon1975}.

\begin{figure}[htp!]
\includegraphics[width=7cm]{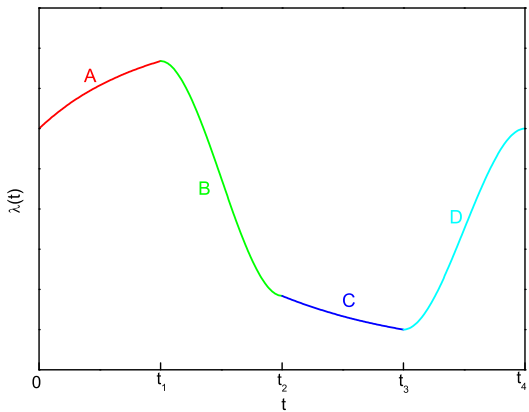}\caption{(Color online) Schematic diagram of protocol (\ref{eq-protctot}). The capital letters A, B, C, and D represent four processes (``isothermal" expansion, ``adiabatic" compression, ``isothermal" compression, and ``adiabatic" expansion) in the thermodynamic cycle, respectively. The times $t_1$, $t_2$, $t_3$, and $t_4$ are not plotted in the same scale.}\label{fig-protocl}
\end{figure}

In addition, we can easily design the protocol for the maximum power according to
Eqs.~(\ref{eq-protadbat}), (\ref{eq-protA}), (\ref{eq-lambda1}), (\ref{eq-lambda22}), (\ref{eq-protC}), (\ref{eq-lambda3}), (\ref{eq-constrat}), (\ref{eq-ch}), and (\ref{eq-cc}). The values of the protocol $\lambda=\lambda(t)$ at times $t_1$, $t_2$, $t_3$, and $t_4$ are found to be $\lambda _{1}=\lambda _{0}\sqrt{(1-c_{h}\tau _{h})/(1-c_{h})}$, $\lambda
_{2}=\lambda _{1}\sqrt{{T_{c}}/{T_{h}}}$, $\lambda _{3}=\lambda _{0}\sqrt{{T_{c}}/{
T_{h}}}$, and $\lambda _{4}=\lambda _{0}$, respectively. The time-dependent protocol for the maximum power may be expressed as
\begin{equation}\label{eq-protctot}
\lambda (t)=\left\{
\begin{array}{l}
\lambda _{0}\sqrt{(1-c_{h}\mathrm{e}^{-2\gamma _{h}t})/(1-c_{h})}
,~0<t\leq t_{1} \\
\lambda _{1}-\lambda _{1}\eta _\mathrm{mP}\Phi\left(\frac{t-t_{1}}{t_{2}-t_{1}}\right) ,~t_{1}<t\leq t_{2} \\
\lambda _{2}\sqrt{[1+c_{c}\mathrm{e}^{-2\gamma _{c}(
t-t_{2}) }]/(1+c_{c})},~t_{2}<t\leq t_{3}\\
\lambda _3+\lambda _{0}\eta _\mathrm{mP}\Phi\left(\frac{t-t_{3}}{t_{4}-t_{3}}\right),~t_{3}<t\leq t_{4}
\end{array}
\right.
\end{equation}
where the function $\Phi(t)$ is still defined as $\Phi(t)\equiv 3t^{2}-2 t^{3}$. The values of $c_h$ and $c_c$ may be calculated from Eqs.~(\ref{eq-ch}) and (\ref{eq-cc}), respectively. The schematic diagram of protocol (\ref{eq-protctot}) is depicted in Fig.~\ref{fig-protocl}.

\section{Summary and discussion\label{sec-summary}}

In this work, we construct a stochastic heat engine by using a time-dependent harmonic potential to control a Brownian particle. By considering the inertial effect of the particle and shortcuts to adiabaticity, we find that the efficiency at maximum power for this microscopically stochastic heat engine is exactly equal to the Curzon-Ahlborn efficiency for endoreversible heat engines.
Our microscopic model has several advantages relative to the Curzon-Ahlborn model. The effective temperatures are well-defined in two ``isothermal" processes. It is unnecessary for us to assume the effective temperatures to be constant as done in the Curzon-Ahlborn model. In particular, Eqs.~(\ref{eq-betaivsA}) and (\ref{eq-betaivsC}) reveal that the effective temperatures are actually inconstant, which is consistent with the fact recently pointed out in Ref.~\cite{AnguloBrown13} that the assumption of constant effective temperatures is not the necessary condition for achieving the Curzon-Ahlborn efficiency. In addition, the finite-time adiabatic processes in our model can be realized with the aid of shortcuts to adiabaticity. It is still unknown how to realize finite-time adiabatic processes for a macroscopic heat engine such as the Curzon-Ahlborn model.

The present stochastic heat engine follows the exquisite model proposed by Schmiedl and Seifert~\cite{Schmiedl2008}. But our start point is different from theirs. In Ref.~\cite{Schmiedl2008}, Schmiedl and Seifert are focused on the overdamping case where the inertial effect of the Brownian particle is neglected while we are concerned with the underdamping case where the inertial effect plays a large role. However, it is this small distinction in the start point that leads to qualitatively different consequences. First, the process in which the protocol $\lambda(t)$ increases (decreases) with time corresponds to a compression (an expansion) of position distribution in Schmiedl-Seifert model, while the process in which the protocol $\lambda(t)$ increases (decreases) with time corresponds to an expansion (a compression) of momentum distribution in our model. In other words, the protocol generating the thermodynamic cycle of a heat engine in Sec.~\ref{sec-model} leads to a refrigerator rather than a heat engine within the framework of Schmiedl-Seifert model. Second, the finite-time adiabatic processes in our model can be realized with the aid of shortcuts to adiabaticity. Suddenly switching the Brownian particle from a hot bath to a cold bath will not produce entropy in the Schmiedl-Seifert model since the position distribution is instantaneously unchanged. However, the mismatch of kinetic energy in this transition will inevitably result in heat exchange between two heat baths. This point is similar to the criticism~\cite{Parrondo96} to Feynman's analysis of the ratchet as a heat engine. That is, the adiabatic transition in the Schmiedl-Seifert model is not genuinely adiabatic. In succeeding work by Seifert's group~\cite{DieterichJSM09}, Schmiedl \emph{et al.} also found that it is important to consider the inertial effect (i.e., the kinetic energy) in the adiabatic transition. They obtained a counter-intuitive result that the minimal work in the adiabatic transition averaged on an initially thermalized ensemble for harmonic potentials is given by the adiabatic work even in the limit of short transition times. It is necessary for us to investigate the relationship between this result and the shortcut to adiabaticity in the future research.

The present model may be generalized in two aspects. First, the reverse thermodynamic cycle will lead to a stochastic refrigerator. The optimization of refrigerators has been investigated by many researchers~\cite{YanChen1990,RocoPRE12,HeChen02,WLTHRpre12,Izumida13}. A reasonable target function is called $\chi$-criterion which is defined as the product of the coefficient of performance of refrigerators and the rate of heat absorbed from the cold bath~\cite{YanChen1990,RocoPRE12}. The coefficient of performance at maximum $\chi$-criterion for endoreversible refrigerators is found to be $\sqrt{T_h/(T_h-T_c)}-1$~\cite{YanChen1990}. It is straightforward to derive the coefficient of performance at maximum $\chi$-criterion for the stochastic refrigerator. Second, the quantum version of the present stochastic heat engine is an intriguing topic. To do that, we need to overcome several difficulties such as the subtle definition of quantum Carnot-like thermodynamic cycle, and the proper definitions of work and heat~\cite{Quanht07,Quanht09}.

Finally, we discuss the realizability of the present stochastic heat engine in experiments. Recently, Blickle and Bechinger~\cite{BechingerNP12} have demonstrated the experimental realization of a microscopically Stirling heat engine by using a time-dependent optical laser trap to control a single colloidal particle of diameter 2.94~$\mu$m. Through qualitative analysis, one can see that the inertial effect for a particle of diameter in micrometers is too small to be detected when we observed in the time-scale of seconds. Therefore, the microscopic heat engine investigated by Blickle and Bechinger is similar to the stochastic heat engine proposed by Schmiedl and Seifert rather than the present model. To enhance the relative strength of inertial effects, one should increase the temporal resolution to microseconds. This difficulty has been overcome by experimental scientists~\cite{Raizen11,Raizen13} who achieved a temporal resolution of $10$~ns for a 1~$\mu$m silica particle. In addition, the generation of shortcuts to adiabaticity mentioned in Sec.~\ref{sec-shtadiab} requires a nonlocal control to the system which might be difficult in experiments. Recently, a local scheme~\cite{Campo100502,DJdCAX14} has been proposed, which can help us overcome the second difficulty. In a word, it is highly promising to realize the present stochastic heat engine in the laboratory.

\section*{Acknowledgement}
The author is grateful for financial supports from the National Natural Science Foundation of China (Grant No. 11322543). He also thanks Adolfo del Campo for his kind discussions on shortcuts to adiabaticity.

\appendix
\section{Detailed derivation of Eq.~(\ref{eq-detaeq})\label{sec-Apdx1}}
The rate of heat absorbed may be defined as
\begin{equation}
\frac{\bar{\mathrm{d}}\left\langle Q\right\rangle }{\mathrm{d}t}=\left\langle \dot{x}\frac{\partial
H}{\partial x}+\dot{p}\frac{\partial H}{\partial p}\right\rangle
\end{equation} where $\bar{\mathrm{d}}$ indicates that the heat is not a state variable and the heat absorbed from the heat bath depends on the detailed process.
The ensemble average $\langle . \rangle$ proceeds in two steps~\cite{Udo08EPJB}. First, we average over
all trajectories which are at time $t$ at given $x$ and $p$ leading to
\begin{equation}
\left\langle \dot{x}|x,p,t\right\rangle =J_{x}/\rho(x,p,t),\left\langle
\dot{p}|x,p,t\right\rangle =J_{p}/\rho(x,p,t),
\end{equation}
where $J_{x}$ and $J_{p}$ are the $x$ and $p$ components of $\mathbf{J}$ in Eq.(\ref{eq-Kramersflux}), respectively. Second, averaging over all $x$ and $p$ with distribution function $\rho(x,p,t)$ leads to
\begin{eqnarray}
\frac{\bar{\mathrm{d}}\left\langle Q\right\rangle }{\mathrm{d}t} &=&\left\langle \dot{x}
\frac{\partial H}{\partial x}+\dot{p}\frac{\partial H}{\partial p}
\right\rangle \nonumber\\
&=&\int dx\int dp \rho \left[  \frac{\partial H}{\partial x}\frac{J_{x}}{\rho}+\frac{\partial H}{\partial p}\frac{J_{p}}{\rho}\right]\nonumber\\
&=&\int dx\int dp\left[  \frac{\partial H}{\partial x}J_{x}+\frac{\partial
H}{\partial p}J_{p}\right]  \nonumber\\
&\equiv &\int dx\int dp\left(\nabla H\right)  \cdot\mathbf{J}.\end{eqnarray}

The above equation may also be derived from the continuity equation of density
\begin{equation}
\frac{\mathrm{d}\rho}{\mathrm{d}t}  =\frac{\partial\rho}{\partial t}
+\nabla\cdot\mathbf{J}\end{equation}
and the continuity equation of energy
\begin{equation}
\frac{\mathrm{d}(H\rho)  }{\mathrm{d}t}   =\frac{\partial(
H\rho)}{\partial t}+\nabla\cdot(  H\mathbf{J}).
\end{equation}

Using the above two equations, we have
\begin{eqnarray}
\frac{\bar{\mathrm{d}}\left\langle Q\right\rangle }{\mathrm{d}t} &=&\left\langle \dot{x}
\frac{\partial H}{\partial x}+\dot{p}\frac{\partial H}{\partial p}
\right\rangle =\left\langle \frac{\mathrm{d}H}{\mathrm{d}t}\right\rangle -\left\langle
\frac{\partial H}{\partial\lambda}\dot{\lambda}\right\rangle\nonumber \\
&=&\int dx\int dp\rho\frac{\mathrm{d}H}{\mathrm{d}t}-\int dx\int dp\rho
\frac{\partial H}{\partial t}\nonumber\\
&=&\int dx\int dp\left\{\left[  \frac{\mathrm{d}(  H\rho)  }{\mathrm{d}
t}-H\frac{\mathrm{d}\rho}{\mathrm{d}t}\right]-\left[
\frac{\partial(  H\rho)  }{\partial t}-H\frac{\partial\rho
}{\partial t}\right] \right\}\nonumber \\
&=&\int dx\int dp\left\{\left[  \frac{\mathrm{d}(  H\rho)  }{\mathrm{d}
t}-\frac{\partial(  H\rho)  }{\partial t}\right]-H\left(  \frac{\mathrm{d}\rho}{\mathrm{d}t}-\frac{\partial\rho}{\partial
t}\right)\right\} \nonumber \\
&=&\int dx\int dp[\nabla\cdot(H\mathbf{J})-H\nabla
\cdot\mathbf{J}]\nonumber \\&=&\int dx\int dp(  \nabla H)  \cdot\mathbf{J}.
\end{eqnarray}
From this equation, we arrive in the first line of Eq.~(\ref{eq-detaeq}). With the consideration of the Hamiltonian (\ref{eq-Hamtion0}), we have $\partial H/\partial x=\partial U/\partial x$ and $\partial H/\partial p=p$. Combining Eq.~(\ref{eq-Kramersflux}), we finally achieve
\begin{eqnarray}
\frac{\bar{\mathrm{d}}\left\langle Q\right\rangle }{\mathrm{d}t} &=&\int dx\int dp\left\{  \frac{\partial U}{\partial x}\left(  p\rho\right)
+p\left[  -\rho\left(  \gamma p+\frac{\partial U}{\partial x}+\frac{\gamma
T}{\rho}\frac{\partial\rho}{\partial p}\right)  \right]  \right\} \nonumber \\
&=&-\int dx\int dp\left[  \gamma p\rho\left(  p+\frac{T}{\rho}\frac
{\partial\rho}{\partial p}\right)  \right].
\end{eqnarray}
The integration of the above equation leads to the second line of Eq.~(\ref{eq-detaeq}).

\section{Detailed derivation of Eq.~(\ref{eq-dHdt})\label{sec-Apdx2}}
From the Hamiltonian (\ref{eq-Hharm}), we have ${\partial H}/{\partial x} =\lambda^{2}x$, ${\partial H}/{\partial p}=p$, and ${\partial H}/{\partial\lambda}=\lambda x^{2}$. Thus
\begin{eqnarray}
\frac{\mathrm{d}H}{\mathrm{d}t}  & =&\frac{\partial H}{\partial x}\dot{x}
+\frac{\partial H}{\partial p}\dot{p}+\frac{\partial H}{\partial\lambda}\dot{\lambda}\nonumber\\
&=&\lambda^{2}x\dot{x} +p\dot{p}  +\lambda x^{2}\dot{\lambda}.\end{eqnarray}
Substituting Eq.~(\ref{eq-Hameq}) into the above equation, we have
\begin{eqnarray}
\frac{\mathrm{d}H}{\mathrm{d}t}
& =&\lambda^{2}x\left(  p-\frac{\dot{\lambda}}{2\lambda}x\right)  +p\left(
-\lambda^{2}x+\frac{\dot{\lambda}}{2\lambda}p\right)  +\lambda x^{2}
\dot{\lambda}\nonumber\\
& =&\frac{\dot{\lambda}}{2\lambda}p^{2}+\frac{\lambda x^{2}\dot{\lambda}}{2}=\frac{\dot{\lambda}}{\lambda}\left(  \frac{p^{2}}{2}+\frac{\lambda^{2}
x^{2}}{2}\right)  \nonumber\\
& =&\frac{\dot{\lambda}}{\lambda}H=H\frac{\mathrm{d}\ln\lambda}{\mathrm{d}t}.
\end{eqnarray}
That is, Eq.~(\ref{eq-dHdt}) holds and its validation is independent of the choice of the protocol $\lambda(t)$.

\end{document}